\documentclass[12pt,preprint]{article}

\usepackage{amsmath}

\usepackage{graphics}

\usepackage{epsfig}

\renewcommand{\vec}[1]{{\bf #1}}

\usepackage{bbm}

\newcommand{\eqb}{\begin{equation}}
\newcommand{\eqe}{\end{equation}}
\newcommand{\dmb}{\begin{displaymath}}
\newcommand{\dme}{\end{displaymath}}

\newcommand{\eab}{\begin{eqnarray}}
\newcommand{\eae}{\end{eqnarray}}

\newcommand{\be}{\begin{equation}}
\newcommand{\ee}{\end{equation}}

\begin{document}

\begin{titlepage}

\begin{flushright} 

\end{flushright}

\vspace{0.6cm}

\begin{center}

\Large{Gap in the black-body spectrum at low temperature}

\vspace{1.5cm}

\large{Markus Schwarz$\mbox{}^\dagger$, Ralf Hofmann$\mbox{}^\dagger$, and Francesco Giacosa$\mbox{}^*$}

\end{center}

\vspace{1.5cm}

\begin{center}

{\em $\mbox{}^\dagger$Institut f\"ur Theoretische Physik\\ 
Universit\"at Heidelberg\\ 
Philosophenweg 16\\ 
69120 Heidelberg, Germany\vspace{0.5cm}\\ 
$\mbox{}^*$Institut f\"ur Theoretische Physik\\ 
Johann Wolfgang Goethe -- Universit\"at\\ 
Max von Laue -- Str. 1\\ 
D-60438 Frankfurt am Main, Germany}

\end{center}

\vspace{1.5cm}

\begin{abstract}

We postulate that the U(1)$_Y$ factor of the Standard Model is an effective 
manifestation of SU(2) gauge dynamics being dynamically broken 
by nonperturbative effects. The modified 
propagation properties of the photon at low temperatures and 
momenta are computed. As a result of strong screening, the presence of a sizable gap in the spectral power of a 
black body at temperatures $T=5\cdots 20\,$K and for low frequencies 
is predicted: A table-top experiment should be able to 
discover this gap. If the gap is observed then 
the Standard Model's mechanism for electroweak symmetry-breaking is endangered by a contradiction 
with Big-Bang nucleosynthesis. Based on our results, we propose an explanation for the stability of 
cold, old, dilute, and large clouds of atomic hydrogen in between 
spiral arms of the outer galaxy.

\end{abstract}

\end{titlepage}

\section{Introduction}

The concept that electromagnetic waves, in analogy to the propagation of distortions 
in fluids, require a medium to travel - the ether -, was employed by Maxwell to 
deduce his famous equations \cite{Maxwell1891}. The ether was 
abandoned with the development of Special Relativity (SR) which is based on two 
empirically founded postulates \cite{Einstein1905}: relativity of uniform motion and constancy 
of the speed of light $c$. SR implies that 
in observing the propagation of a monochromatic light wave (photon) 
no inertial frame of reference is singled out. This situation 
changes when thermalized radiation is considered: 
The very process of thermalization proceeds by interactions between the photon and electrically charged, 
massive matter. The latter's center of inertia, however,  
defines a preferred rest frame. This, at least, is the standard notion of 
how a temperature $T$ emerges in a photon gas. 

On a microscopic level, photon interactions are described by quantum mechanical 
transitions. The underlying and very successful field theory, Quantum Electrodynamics (QED) 
\cite{Tomonaga1946,Schwinger1949,Feynman1949}, is based on a U(1) gauge group. 
In the present Standard Model
of particle physics (SM) a progenitor of this symmetry is 
called U(1)$_Y$. On a thermodynamical level, the microscopic details of the 
emission and absorption processes are averaged away. 
As a consequence, thermalized photons exhibit a 
universal black-body spectrum whose shape solely 
depends on $T$ and the two constants 
of nature $c$ and $\hbar$. 

In this Letter we explore the possibility that the U(1)$_Y$ factor 
of the SM's gauge group is not fundamental \cite{Hofmann2005,BruntKnee2001,WMAP2003,PVLAS}. In this context, 
QED may break down under exceptional conditions. 
Such an exception would take place in a thermalized photon gas at 
temperatures not far above 2.73\,Kelvin (K): The notion of a 
gas of interaction-free photons then would require revision. Namely, 
embedding U(1)$_Y$ into the fundamental, nonabelian gauge 
group SU(2), invokes a mass scale 
$\Lambda\sim 10^{-4}$ electron volts (eV) 
\cite{Hofmann2005,Hofmann20051,GiacosaHofmann2005,SchwarzHofmannGiacosa2006} 
which affects black-body spectra. 
The SU(2) gauge symmetry implies the existence and relevance of 
the Yang-Mills scale $\Lambda$ on the quantum level \cite{Politzer1973,GrossWilczek1973}. We will 
discuss below why we use the name SU(2)$_{\tiny\mbox{CMB}}$ (CMB for 
{\sl Cosmic Microwave Background}) for the fundamental gauge symmetry also 
describing photon propagation at low temperatures.  

Let us now list some bulk properties of 
SU(2) gauge theory in four dimensions \cite{Hofmann2005,SchwarzHofmannGiacosa2006,HS1977,Nahm1984,
LeeLu1998,KraanVanBaalNPB1998,vanBaalKraalPLB1998,Diakonov2004,
HerbstHofmann2004,HerbstHofmannRohrer2004}. First, three phases exist: 
a deconfining, a preconfining, and a confining 
one (in order of decreasing temperature). Only the deconfining phase is relevant for the present 
discussion. Upon a spatial coarse-graining this phase is described by an effective field 
theory \cite{Hofmann2005}. Second, for $T$ much greater than the 
scale $\Lambda$ all thermodynamical quantities reach their Stefan-Boltzmann 
limits in a power-like way. Third, there is a nontrivial 
ground state, obeying an equation of state $P_{gs}=-\rho_{gs}=-4\pi\Lambda^3 T$, 
in the deconfining phase. This ground state is tied to the presence of 
interacting topological defects (calorons, topology-changing quantum fluctuations 
\cite{Hofmann2005,HerbstHofmann2004}). Through interactions with the ground state 
two types of gauge modes ($V^\pm$) acquire a temperature-dependent mass while the 
third type remains massless ($\gamma$). It is important to 
note that at a critical temperature $T_c$ (boundary between deconfining and preconfining phase) 
$\gamma$'s partners $V^\pm$ acquire an infinite mass and thus decouple thermodynamically. Moreover, within the
deconfining phase quantum fluctuations are severely constrained 
in the effective theory: The interactions of the three types of gauge bosons 
are very weak \cite{SchwarzHofmannGiacosa2006,HerbstHofmannRohrer2004}. 
 
Taking these interactions 
into account, $\gamma$'s dispersion law 
\eqb
\label{NoSC}
\omega^2=\vec{p}^2 
\eqe
modifies as \cite{SchwarzHofmannGiacosa2006}
\eqb
\label{ASDR}
\omega^2=\vec{p}^2+G(\omega,\vec{p},T,\Lambda)\,.
\eqe
In Eq.\,(\ref{ASDR}) $\omega$ denotes the energy of a $\gamma$-mode with 
spatial momentum $\vec{p}$. The screening function $G$ depends on $\omega,\,\vec{p}$, 
temperature $T$, and the Yang-Mills scale $\Lambda$. Notice that in writing Eqs.\,(\ref{NoSC}) and 
(\ref{ASDR}) we use 
natural units: $c=\hbar=k_B=1$ where $k_B$ is Boltzmann's constant. Our results indicate 
that for $T\gg T_c$ the function $G$ is negative with a negligibly 
small modulus (antiscreening). However,$T$ a few times $T_c$ and for small 
$|\vec{p}|$ the function $G$ becomes positive and 
reaches sizable values ($>|\vec{p}|^2$). That is, the $\gamma$-mode acquires a screening mass. 
If emitted with the dispersion law of 
Eq.\,(\ref{NoSC}) then the dispersion law of Eq.\,(\ref{ASDR}) 
is violated: $\gamma$ can penetrate the plasma only up to a 
distance $\sim G^{-1/2}$. A useful analogy is a rain-drop falling 
onto the surface of a lake where it is absorbed immediately. 

At $T_{c}=\frac{\lambda_c}{2\pi}\Lambda$ ($\lambda_c=13.87$ \cite{Hofmann2005}) 
for the (second-order like) transition to the
preconfining phase we have 
\eqb
\label{Gcrit}
\lim_{T\searrow T_c}\,G(\omega,\vec{p},T,\Lambda)=0\,.
\eqe
Thus no (anti)screening is seen in $\gamma$-propagation at $T=T_c$. The above results match with observational ($T\sim
T_{\tiny\mbox{CMB}}$) and daily ($T\gg T_{\tiny\mbox{CMB}}$) experience that the photon's dispersion law is the one in 
Eq.\,(\ref{NoSC}). Hence we are led to identify $\gamma$ with the photon 
(U(1)$_Y\subset$ SU(2)$_{\tiny\mbox{CMB}}$) and 
conclude that $T_c=T_{\tiny\mbox{CMB}}\sim 2.73\,\mbox{K}=2.35\times 10^{-4}\,$eV 
for SU(2)$_{\tiny\mbox{CMB}}$. In spite of the fact that 
such an identification is rather unconventional we believe that it is 
worthwhile to pursue its consequences, be it only to falsify such a scenario. 

Within a cosmological context one derives 
that the photon remains massless only for a finite period of time, 
$\Delta t\le 2\,$billion years, in the future \cite{GiacosaHofmann2005}. This 
is due to the fact that the transition to the 
preconfining phase invokes a nonvanishing coupling of the photon 
to a newly emerging ground state: The remaining gauge symmetry 
U(1)$_Y$ is then broken dynamically (in contrast to the 
deconfining phase); a familiar effect in macroscopic 
superconductivity \cite{LandauGinzburg1950,Abrikosov1956}.

\section{Modified black-body spectra at low temperatures and low frequencies\label{BBSpec}}

In \cite{SchwarzHofmannGiacosa2006} we have calculated the function $G$ appearing 
in Eq.\,(\ref{ASDR}) from the photon's polarization tensor. 
We have made the assumption that $\omega=|\vec{p}|$ on the right-hand side 
of Eq.\,(\ref{ASDR}). (For $\omega=|\vec{p}|$ $G$ is {\sl real}.) 
This is a consistent approximation as long as $\frac{|G|}{\omega^2}\ll 1$, see below. 
In Figs.\,\ref{Fig-1} and \ref{Fig-2} the function $\log_{10}\left|\frac{G}{T^2}\right|$ 
is depicted in dependence of (dimensionless) temperature $\lambda\equiv\frac{2\pi T}{\Lambda}$ 
and (dimensionless) momentum $X\equiv\frac{|\vec{p}|}{T}$, respectively.      
\begin{figure}
\begin{center}
\leavevmode
\leavevmode
\vspace{4.3cm}
\includegraphics{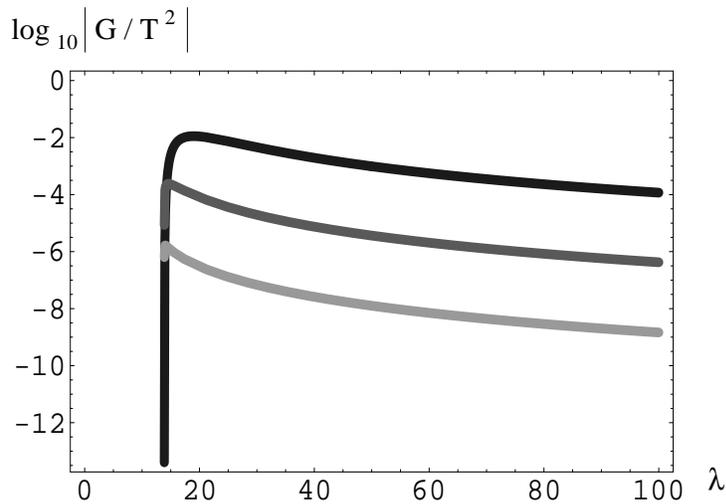}
\end{center}
\caption{\protect{\label{Fig-1}}$\log_{10}\left|\frac{G}{T^2}\right|$ as a function of 
$\lambda\equiv\frac{2\pi T}{\Lambda}$ for $X=1$ (black), $X=5$ (dark grey), and $X=10$ 
(light grey) where $X\equiv\frac{|\vec{p}|}{T}$.}      
\end{figure}
\begin{figure}
\begin{center}
\leavevmode
\leavevmode
\vspace{5.9cm}
\includegraphics{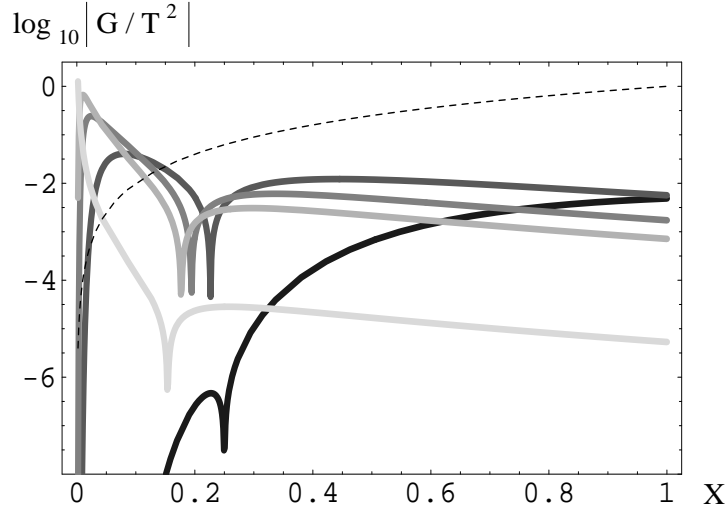}
\end{center}
\caption{\protect{\label{Fig-2}}$\log_{10}\left|\frac{G}{T^2}\right|$ as a function of $X$ 
for $\lambda=1.12\,\lambda_{c}$ (black), $\lambda=2\,\lambda_{c}$ (dark grey), 
$\lambda=3\,\lambda_{c}$ (grey), $\lambda=4\,\lambda_{c,E}$ (light grey), $\lambda=20\,\lambda_{c}$ 
(very light grey). The dashed curve is a 
plot of the function $f(X)=2\log_{10}X$. Photons are strongly screened at 
$X$-values for which $\log_{10}\left|\frac{G}{T^2}\right|>f(X)$, that is, to the left 
of the dashed line. The dips correspond to the zeros of $G$.}      
\end{figure}

As Fig.\,\ref{Fig-1} shows, at fixed values of $X$ 
the function $|G|$ falls off in a power-like way at large 
temperatures. Equidistance of the curves for equidistant values of 
$X\ge 1$ indicates exponential suppression in $X$. 
For $T\searrow T_{c}$ the thermodynamical decoupling of 
$V^{\pm}$-modes at the phase boundary leads to a 
rapid drop of $|G|$. In Fig.\,\ref{Fig-2} the 
low-momentum behavior of $|G|$ at fixed temperatures 
not far above $T_c$ is depicted. For SU(2)$_{\tiny\mbox{CMB}}$ the (dimensionless) 
temperatures $\lambda=1.12\,\lambda_{c},\,2\,\lambda_{c},\,3\,\lambda_{c},\,4\,\lambda_{c,E}\,$ 
and $\lambda=20\,\lambda_{c}$ convert into $T=3.02\,\mbox{K},\,5.5\,\,\mbox{K},
\,8.2\,\,\mbox{K},\,10.9\,\,\mbox{K}$, and $T=55\,\mbox{K}$, 
respectively. For $T=3.02\,\mbox{K}\sim T_{\tiny\mbox{CMB}}$ (black curve) 
and $T=55\,\mbox{K}$ (very light grey curve) 
the regime, where photons are strongly screened, 
is too small to be resolved by existing low-temperature black-body (LTBB) observations and 
experiments \cite{Morozova1993}. For the other temperatures considered in Fig.\,\ref{Fig-2} 
there is a sizable range of $X$-values for this effect. (We only mention here 
that photons do propagate again at very small momenta \cite{SchwarzHofmannGiacosa2006}.) 
\begin{figure}
\begin{center}
\leavevmode
\leavevmode
\vspace{5.8cm}
\includegraphics{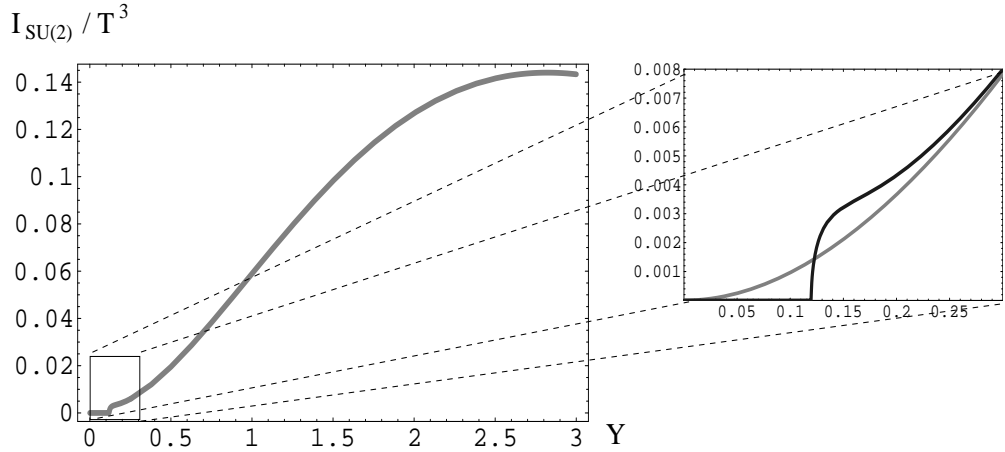}
\end{center}
\caption{\protect{\label{Fig-3}}Dimensionless black-body spectral power 
$\frac{I_{\tiny\mbox{SU(2)}}}{T^3}$ as a function of the dimensionless frequency 
$Y\equiv\frac{\omega}{T}$. The black curve in the magnified region depicts the modification of 
the spectrum as compared to $\frac{I_{\tiny\mbox{U(1)}}}{T^3}$ (grey curve) 
for $T=10\,$K.}      
\end{figure}

The spectral power $I_{\tiny\mbox{U(1)}}(\omega)$ for a black body subject to the 
gauge symmetry U(1) is given as
\eqb
\label{PiBB}
I_{\tiny\mbox{U(1)}}(\omega)=\frac{1}{\pi^2}\,\frac{\omega^3}{\exp[\frac{\omega}{T}]-1}\,.
\eqe
For SU(2)$_{\tiny\mbox{CMB}}$ this modifies as
\eqb
\label{PmBB}
I_{\tiny\mbox{U(1)}}(\omega)\to I_{\tiny\mbox{SU(2)}}(\omega)=I_{\tiny\mbox{U(1)}}(\omega)\times
\frac{\left(\omega-\frac{1}{2}\frac{d}{d\omega}G\right)\sqrt{\omega^2-G}}{\omega^2}\,
\theta(\omega-\omega^*)\,
\eqe
where $\omega^*$ is the root of $\omega^2=G$, and $\theta(x)$ is the Heaviside step 
function. In Fig.\,\ref{Fig-3} the modification of the black-body spectrum according 
to Eq.\,(\ref{PmBB}) is depicted for $T=10\,$K: There is no spectral 
power at frequencies $\omega<0.12\,T$ whereas there is a (rapidly decreasing) 
excess at frequencies $\omega>0.12\,T$. 

\begin{figure}
\begin{center}
\leavevmode
\leavevmode
\vspace{5.8cm}
\includegraphics{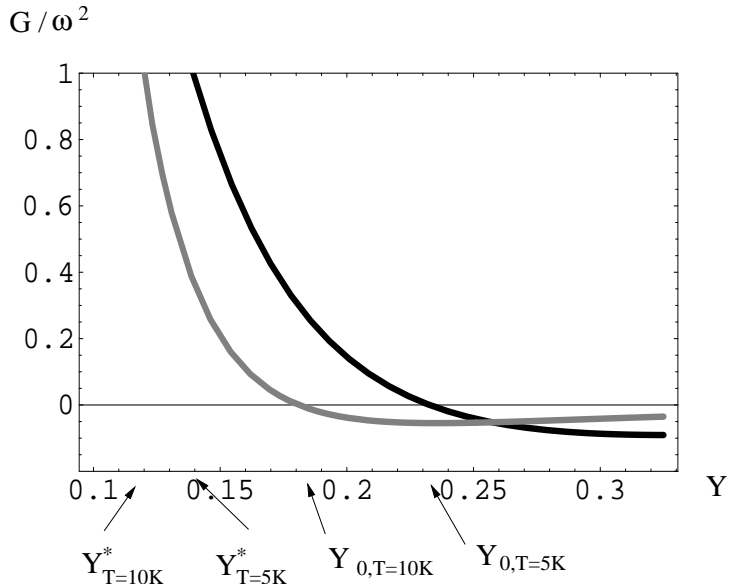}
\end{center}
\caption{\protect{\label{Fig-4}}$\frac{G}{\omega^2}$ as a function 
of $Y\equiv\frac{\omega}{T}$ for $T=5\,$K (black curve) and 
$T=10\,$K (grey curve).}      
\end{figure}

Let us now investigate how reliable the approximation 
$\omega=|\vec{p}|$ is when evaluating the function $G$. In 
Fig.\,\ref{Fig-4} a plot of $\frac{G}{\omega^2}$ is shown as a function 
of $Y\equiv\frac{\omega}{T}$ for $T=5\,$K (black curve) and 
for $T=10\,$K (grey curve) in the vicinity of $G$'s zero $Y_0$. To the 
right of $Y_0$ the condition $\frac{|G|}{\omega^2}\ll 1$ is well 
satisfied for $\omega=|\vec{p}|$, to the left of $Y_0$ this continues to be 
a reasonable approximation almost down to $Y^*\equiv\frac{\omega^*}{T}$ because of 
the large negative slope of the function $\frac{G}{\omega^2}$ in the vicinity of $Y^*$: 
Although our approximation is doomed to break 
down at $Y^*$ it is still valid for values of $Y$ slightly above $Y^*$ where the 
tendency towards large $G$ is seen. For an experiment to 
detect the reshuffling of spectral power as indicated in Fig.\,\ref{Fig-3} 
the spectrometer must not be further away from the aperture of the LTBB than $G^{-1/2}$.

Let us now discuss how sensitive the measurement of the LTBB spectral intensity 
$I_{\tiny\mbox{SU(2)}}(\omega)$ needs to 
be in order to detect the spectral gap setting in at $\omega^*$. 
A useful criterion is determined by the ratio $R(Y^*)$ of $I_{\tiny\mbox{U(1)}}(\omega^*)$ and 
$I_{\tiny\mbox{U(1)}}(\omega_{\tiny\mbox{max}})$ where $\omega_{\tiny\mbox{max}}=2.82\,T$ 
(in natural units) is the position of the maximum of $I_{\tiny\mbox{U(1)}}$ :
\eqb
\label{R(xt)}
R(Y^*)\equiv\frac{I_{\tiny\mbox{U(1)}}(\omega^*)}{I_{\tiny\mbox{U(1)}}
(\omega_{\tiny\mbox{max}})}=\frac{1}{1.42144}\,\,\frac{(Y^*)^3}{\exp(Y^*)-1}\,.
\eqe
For $T=80\,$K, which was experimentally realized in \cite{Morozova1993}, 
we have $R(Y^*=0.0366)=9\times 10^{-4}$. To achieve such a high precision is a 
challenging task. To the best of the authors knowledge only the overall and not the spectral 
intensity of the LTBB was measured in 
\cite{Morozova1993}. For $T=5\,$K one has $R(Y^*=0.14)=1.2\times 10^{-2}$. 
Thus at low temperatures the precision required to detect the spectral gap is within the 
1\%-range. It is, however, experimentally challenging to cool 
the LTBB down to these low temperatures. To the best of the authors 
knowledge a precision measurement of the spectral power in the 
low-frequency regime of a LTBB at $T=5\cdots 10\,$K has not yet been performed. We 
know, however, that such an experiment is well feasible \cite{Sapritsky}: It will 
represent an important and inexpensive (on particle-physics scales) test
of the postulate SU(2)$_{\tiny\mbox{CMB}}\stackrel{\tiny\mbox{today}}=$U(1)$_Y$.

\section{Stability of dilute and cold hydrogen clouds in 
the outer galaxy\label{SHC}}

In \cite{BruntKnee2001,Dickey2001} the existence of a large (up to 2\,kpc), 
old (estimated age $\sim$ 50 million years), 
cold (mean brightness temperature $T_B\sim 20\,$K with cold regions of $T_B\sim 5\cdots 10\,$K), 
dilute (number density: $\sim 1.5$\,cm$^{-3}$) and massive ($1.9\times 10^7$ solar masses) 
innergalactic cloud (GSH139-03-69) of atomic hydrogen (HI) forming an 
arc-like structure in between spiral arms 
was reported. In \cite{Dickey} and references therein smaller structures of this type were identified. 
These are puzzling results which do not fit into the dominant model 
for the interstellar medium \cite{Dickey2001}. Moreover, considering the typical 
time scale for the formation of H$_2$ molecules out of HI of about $10^{6}\,$yr 
\cite{Dickey} at these low temperatures and low densities clashes with the inferred age 
of the structure observed in \cite{BruntKnee2001}.

To the best of our knowledge there is no standard 
explanation for the existence and the stability 
of such structures. We wish to propose a scenario possibly 
explaining the stability based on SU(2)$_{\tiny\mbox{CMB}}$. 
Namely, at temperatures $T_B\sim 5\cdots 10\,$K, corresponding to $T_B\sim 2\cdots 4\,T_{\tiny\mbox{CMB}}$, 
the function $G$ for photons with momenta ranging 
between $|\vec{p}^*|\sim 0.15\,T_B>|\vec{p}_c|>|\vec{p}_{\tiny\mbox{low}}|$ is such 
that it strongly suppresses their propagation, see Figs.\,\ref{Fig-2}. 
We mention in passing only that $|\vec{p}_{\tiny\mbox{low}}|<0.02\,T_B$ depends rather 
strongly on temperature \cite{SchwarzHofmannGiacosa2006}.    

Incidentally, the regime for the wavelength $l_c$ associated with $|\vec{p}_c|$ is 
comparable to the interatomic distance $\sim 1\,$cm in GSH139-03-69: 
At $T=5\,$K we have $l^*=2.1\,\mbox{cm}\le l_c\le 8.8\,\mbox{cm}=l_{\tiny\mbox{low}}$, at 
$T=10\,$K we have $l^*=1.2\,\mbox{cm}\le l_c\le 1.01\,\mbox{m}=l_{\tiny\mbox{low}}$. Thus the 
photons mediating the dipole interaction between HI 
particles practically do not propagate: the dipole force at these distances appears 
to be switched off. As a consequence, H$_2$ molecules are prevented from 
forming at the temperatures and densities which are typical for GSH139-03-69. 

The astrophysical origin of the structure GSH139-03-69 appears 
to be a mystery. The point we are able to make here is that 
once such a cloud of HI particles has formed it likely 
remains in this state for a long period of time.  

\section{Implications for the Standard Model and 
Conclusions}

We will now argue that if SU(2)$_{\tiny\mbox{CMB}}$ is, indeed, 
realized in Nature then the SM's Higgs mechanism for electroweak symmetry-breaking is not. 
The key is to work out the consequences of SU(2)$_{\tiny\mbox{CMB}}$ for Big-Bang nucleosynthesis. 
The SM predicts that within this cosmological epoch the number of relativistic 
degrees of freedom $g_{\ast }$ is given as 
\begin{equation}
g_{\ast }=5.5+\frac{7}{4}N_{\nu }  \label{reldof}
\end{equation}%
where $1.8\leq N_{\nu }\leq 4.5$ \cite{Eidelman2005}. 
This prediction relies on
the following consideration: the neutron to proton fraction $n/p$ at freeze-out
is given as $n/p=\exp [-Q/T_{{\tiny \mbox{fr}}}]\sim 1/6$ where $Q=1.293\,$%
\thinspace MeV is the neutron-proton mass difference, and one has 
\begin{equation}
T_{{\tiny \mbox{fr}}}\sim \left( \frac{g_{\ast }G_{N}}{G_{F}^{4}}\right)
^{1/6}\,.  \label{Tfreeze}
\end{equation}%
In Eq.\thinspace (\ref{Tfreeze}) $G_{N}$ denotes Newton's constant, and 
\eqb
\label{GF}
G_{F}=\pi \frac{\alpha _{w}}{\sqrt{2}\,m_{W}^{2}} \sim 1.17\times 10^{-5}\,\mbox{GeV}^{-2}
\eqe
is the Fermi coupling at zero temperature. To use the zero-temperature value of 
$G_{F}$ at $T_{{\tiny \mbox{fr}}}=1$\,MeV, as it is done in Eq.\,(\ref{Tfreeze}), 
is well justified by the large `electroweak scale' $v=247$\thinspace GeV in the SM: the vacuum 
expectation of the Higgs-field. Invoking 
SU(2)$_{\tiny\mbox{CMB}}$ yields additional six relativistic degrees of
freedom ($V^\pm$ with three polarizations each) 
at $T_{{\tiny \mbox{fr}}}=1\,$MeV: a result which exceeds the above cited 
upper bound for $N_{\nu }$. The value of $T_{{\tiny \mbox{fr}}}\sim 1$\,MeV is rather reliably 
extracted from the primordial $^4$He abundance $Y_p\sim 0.25$ and the 
subsequent determination of $n/p\sim 1/6$ \cite{Eidelman2005}. To save this value of $T_{{\tiny \mbox{fr}}}$ 
one needs to prescribe a value 
of $G_{F}$ at $T_{{\tiny \mbox{fr}}}=1$\,MeV which is about 12\,\% larger 
than the value of $G_{F}$ in Eq.\,(\ref{GF}). Since $G_F$ at $T=0$ is measured 
to per mille accuracy there would be a contradiction with electroweak SM physics. A larger value of 
$G_F(T=T_{\tiny\mbox{fr}})$ is, however, expected if the weak interactions are based on a higgsless SU(2) gauge theory of 
Yang-Mills scale $\Lambda\sim 0.5\,$MeV, see the 
discussion in \cite{GiacosaHofmann2005}.  

The prime physical system, for which our results are relevant, is the cosmic 
microwave background. Namely, at small redshift ($z<20$) the screening 
effects of intermediate $V^\pm$ bosons have the potential to explain the large-angle 
anomalies in the power spectra as they were reported in \cite{WMAP2003}. 

For the above reasons the importance of an experimental verification or falsification 
of the postulate SU(2)$_{\tiny\mbox{CMB}}\stackrel{\tiny\mbox{today}}=$U(1)$_Y$ 
involving the low-temperature, low-frequency regimes in 
black-body spectra is evident.

\section*{Acknowledgments}

F. G. acknowledges financial support by the Virtual Institute VH-VI-041
"Dense Hadronic matter \& QCD phase transitions" of the Helmholtz
association.

\baselineskip25pt

\end{document}